\documentclass[aps,prd,nofootinbib,preprint]{revtex4}

\usepackage{graphicx}
\usepackage{psfrag}

\begin{document}
\def\dirac#1{#1\llap{/}}
\def\pv#1{\mathbf{#1}_\perp}

\title{Consistent Analysis of the $B\to\pi$ Transition
Form Factor in the Whole Physical Region}
\author{Tao, Huang$^{1,2}$\footnote{email:
huangtao@mail.ihep.ac.cn} and Xing-Gang, Wu$^{2}$\footnote{email:
wuxg@mail.ihep.ac.cn}}
\address{$^1$CCAST(World
Laboratory), P.O.Box 8730, Beijing 100080, P.R.China\\
$^2$Institute of High Energy Physics, Chinese Academy of Sciences,
P.O.Box 918(4), Beijing 100049, P.R. China\footnote{Mailing
address}}

\begin{abstract}
In the paper, we show that the $B\to\pi$ transition form factor
can be calculated by using the different approach in the different
$q^2$ regions and they are consistent with each other in the whole
physical region. For the $B\to\pi$ transition form factor in the
large recoil regions, one can apply the PQCD approach, where the
transverse momentum dependence for both the hard scattering part
and the non-perturbative wavefunction, the Sudakov effects and the
threshold effects are included to regulate the endpoint
singularity and to derive a more reliable PQCD result. Pionic
twist-3 contributions are carefully studied with a better endpoint
behavior wavefunction for $\Psi_p$ and we find that its
contribution is less than the leading twist contribution. Both the
two wavefunctions $\Psi_B$ and $\bar\Psi_B$ of the B meson can
give sizable contributions to the $B\to\pi$ transition form factor
and should be kept for a better understanding of the B decays. The
present obtained PQCD results can match with both the QCD
light-cone sum rule results and the extrapolated
lattice QCD results in the large recoil regions.\\


\end{abstract}
\maketitle

\section{Introduction}

There are various approaches to calculate the $B\to\pi$ transition
form factor, such as the lattice QCD
technique\cite{lattice,lattice2,lattice3}, the QCD light-cone sum
rules (LCSRs)\cite{sumhuang,sumrule,sumrule2,pball} and the
perturbative QCD (PQCD)
approach\cite{wirbel,huangl,lihn,lihn1,weiy2,lucai}. The PQCD
calculation is reliable only when the involved energy scale is
hard enough, i.e. in the large recoil regions. Due to the
restriction to the $\pi$ energies smaller than the inverse lattice
spacing, the lattice QCD calculation becomes more difficult in the
large recoil regions and at the present, the lattice QCD results
of the $B\to\pi$ transition form factor are available only for
soft regions, i.e. $q^2>15GeV^2$. The lattice QCD results can be
extrapolated to small $q^2$ regions, and the different
extrapolation methods might cause uncertainties about
$5\%$\cite{lattice2}. While, the QCD LCSRs can involve both the
hard and the soft contributions below $q^2<18GeV^2$\cite{sumhuang}
and can be extrapolated to higher $q^2$
regions\cite{sumrule,sumrule2,pball}. Therefore, the results from
the PQCD approach, the lattice QCD approach and the QCD LCSRs are
complementary to each other, and by combining the results from
these three methods, one may obtain a full understanding of the
$B\to\pi$ transition form factor in its physical region, $0\leq
q^2 \leq (M_B-M_\pi)^2\simeq 25GeV^2$.

Certain exclusive process involving hadrons can be described by
PQCD if the momentum transfer is sufficiently large. The amplitude
can be factorized into the convolution of the non-perturbative
wavefunction for each of the hadrons with a PQCD calculable
hard-scattering amplitude. The PQCD factorization theorem has been
worked out in Refs.\cite{li,liyu} based on the earlier works on
the applications of PQCD to hard exclusive processes \cite{hard}.
In the present paper, we shall use the PQCD approach to calculate
the $B\to\pi$ transition form factor in the large recoil regions.

In the PQCD approach based on collinear factorization theorem, a
direct calculation of the one-gluon-exchange diagram for the $B$
meson transition form factor suffers singularities from the
endpoint region of a momentum fraction $x\to 0$. Because of these
singularities, it was claimed that $B\to\pi$ transition form
factor is dominated by soft dynamics and not calculable in
PQCD\cite{pqcdno}. In fact, in the endpoint region the parton
transverse momenta $\mathbf{k}_\perp$ are not negligible. After
including the parton transverse momenta, large double logarithmic
corrections $\alpha_s \ln^2 k_\perp$ appear in higher order
radiative corrections and must be summed to all orders. In
addition, there are also large logarithms $\alpha_s \ln^2x$ which
should also be summed (threshold resummation\cite{threshold1}).
The relevant Sudakov form factors from both $k_\perp$ and the
threshold resummation can cure the endpoint singularity which
makes the calculation of the hard amplitudes infrared safe, and
then the main contribution comes from the perturbative regions.

An important issue for calculating the $B\to\pi$ transition form
factor is whether we need to take both the two wavefunctions
$\Psi_B$ and $\bar\Psi_B$ into consideration or simply $\Psi_B$ is
enough? In literature, many authors (see
Refs.\cite{huangl,lihn,lihn1}) did the phenomenological analysis
with only $\Psi_B$, setting $\bar\Psi_B=0$ (or strictly speaking,
ignoring the contributions from $\bar\Psi_B$). However, As has
been argued in Refs.\cite{wu1,descotes}, one may observe that the
distribution amplitudes (DAs) of those two wavefunctions have a
quite different endpoint behavior, such difference may be strongly
enhanced by the hard scattering kernel. Even though $\bar\Psi_B$
(with the definition in Ref.\cite{lucai}) is of subleading order
contribution, there is no convincing motivation for setting
$\bar\Psi_B=0$. In the present paper, we shall keep both the two
wavefunctions $\Psi_B$ and $\bar\Psi_B$ to do our calculations and
show to what extent the $\bar\Psi_B$ can affect the final results.
Another issue we need to be more careful is about the pionic
twist-3 contributions. Based on the asymptotic behavior of the
twist-3 DAs, especially $\phi^{as}_p(x)\equiv 1$, most of the
people pointed out a large twist-3 contribution\cite{li1,weiy2} to
the $B\to\pi$ transition form factor, i.e. bigger than that of the
leading twist in almost all of the energy regions. In
Ref.\cite{huangwu}, the authors have made a detailed analysis on
the model dependence of the twist-3 contributions to the pion
electro-magnetic form factor, and have raised a new twist-3
wavefunction with a better endpoint behavior for $\Psi_p$, which
is derived from the QCD sum rule moment calculation\cite{huang3}.
And their results show that with such new form for $\Psi_p$, the
twist-3 contributions to the pion electro-magnetic form factor are
power suppressed in comparison to the leading twist contributions.
According to the power counting rules in Ref.\cite{li1}, the
pionic twist-2 and twist-3 contributions should be of the same
order for the case of the B meson decays. With the new form for
$\Psi_p$\cite{huangwu}, we show that for the case of the $B\to\pi$
transition form factor, even though the twist-3 contributions are
of the same order of the leading twist contributions, its values
are less than the leading twist contribution.

The purpose of the paper is to examine the $B\to\pi$ transition
form factor in the PQCD approach, and to show how the PQCD results
can match with the QCD LCSR results and the extrapolated lattice
QCD results. In the PQCD approach, the full transverse momentum
dependence ($k_T$-dependence) for both the hard scattering part
and the non-perturbative wavefunction, the Sudakov effects and the
threshold effects are included to cure the endpoint singularity.
In section II, based on the $k_T$ factorization formulism, we give
the PQCD formulae for the $B\to\pi$ transition form factor in the
large recoil regions. In section III, we give our numerical
results and carefully study the contributions from $\Psi_B$ and
$\bar\Psi_B$, and those from the different pionic twist
structures. The slope of the obtained form factors
$F_{+,\;0}^{B\pi}(q^2)$ in the large recoil regions can match with
those obtained from other approaches. Conclusion and a brief
summary are presented in the final section.

\section{$B\to\pi$ transition form factor in the large recoil regions}

First, we give our convention on the kinematics. For convenience,
all the momenta are described in terms of the light cone (LC)
variables. In the LC coordinate, the momentum is described in the
form, $k=(\frac{k^+}{\sqrt 2}, \frac{k^-}{\sqrt 2},
\mathbf{k}_\bot)$, with $k^{\pm}=k^0\pm k^3$ and
$\mathbf{k}_\bot=(k^1, k^2)$. The scalar product of two arbitrary
vectors $A$ and $B$ is, $A\cdot B= \frac{A^+B^- + A^-B^+}{ 2} -
\mathbf{A}_\bot\cdot\mathbf{B}_\bot$. The pion mass is neglected
and its momentum is chosen to be in the minus direction. Under the
above convention, we have $P_B =\frac{M_B}{\sqrt
2}(1,1,\mathbf{0}_\perp)$, $P_{\pi}=\frac{M_B}{\sqrt 2}(0,\eta
,\mathbf{0}_\bot)$ and $\bar{P}_{\pi}=\frac{M_B}{\sqrt
2}(\eta,0,\mathbf{0}_\bot)$, with $\eta=1-\frac{q^2}{M_B^2}$ and
$q=P_B-P_{\pi}$.

The two $B\to \pi$ transition form factors $F_+^{B\pi}(q^2)$ and
$F_0^{B\pi}(q^2)$ are defined as follows:
\begin{equation}\label{eq:bpi1}
\langle \pi(P_{\pi})|\bar u \gamma_{\mu}b|\bar
B(P_B)\rangle=\left((P_B+P_{\pi})_{\mu}-
\frac{M_B^2-m_{\pi}^2}{q^2}q_{\mu}\right)F_+^{B\pi}(q^2)+
 \frac{M_B^2-m_{\pi}^2}{q^2} q_{\mu}F_0^{B\pi}(q^2),
\end{equation}
where $F_+^{B\pi}(0)$ should be equal to $F_0^{B\pi}(0)$ so as to
cancel the poles at $q^2=0$.

The amplitude for the $B\to \pi$ transition form factor can be
factorized into the convolution of the wavefunctions for the
respective hadrons with the hard-scattering amplitude. The
wavefunctions are non-perturbative and universal. The momentum
projection for the matrix element of the pion has the following
form,
\begin{equation}
M_{\alpha\beta}^{\pi} = \frac{i f_{\pi}}{4} \Bigg\{
\slash\!\!\!p\,\gamma_5\,\Psi_{\pi}(x, \mathbf{k_\perp})-
m^p_0\gamma_5 \left(\Psi_p(x, \mathbf{k_\perp})
-i\sigma_{\mu\nu}\left(n^{\mu}\bar{n}^{\nu}\,\frac{\Psi_{\sigma}'(x,
\mathbf{k_\perp})}{6}-p^\mu\,\frac{\Psi_\sigma(x,
\mathbf{k_\perp})}{6}\, \frac{\partial}{\partial
\mathbf{k}_{\perp\nu}} \right)\right)
\Bigg\}_{\alpha\beta},\label{benek}
\end{equation}
where $f_{\pi}$ is the pion decay constant and $m^p_0$ is the
parameter that can be determined by QCD sum rules\cite{huang3}.
$\Psi_{\pi}(x,\mathbf{k_{\perp}})$ is the leading twist (twist-2)
wave function, $\Psi_p(x,\mathbf{k_{\perp}})$ and
$\Psi_{\sigma}(x,\mathbf{k_{\perp}})$ are sub-leading twist
(twist-3) wave functions, respectively. $\Psi_{\sigma}'(x,
\mathbf{k_\perp})=\partial \Psi_{\sigma} (x,
\mathbf{k_\perp})/\partial x$, $n=(\sqrt{2},0,\mathbf{0}_\bot)$
and $\bar{n}=(0,\sqrt{2},\mathbf{0}_\bot)$ are two null vectors
that point to the plus and the minus directions, respectively. The
momentum projection for the matrix element of the B meson can be
written as \cite{weiy2,BenekeFeldmann}:
\begin{equation}\label{projectorB}
M^B_{\alpha\beta}=-\frac{if_B}{4}
\left\{\frac{\dirac{p}_B+M_B}{2}\left[\dirac{n}
\Psi^+_B(\xi,\mathbf{l_\bot}) +
\bar\dirac{n}\Psi^-_B(\xi,\mathbf{l_\bot})-\Delta(\xi,
\mathbf{l}_{\bot}) \gamma^\mu \frac{\partial} {\partial
l_\perp^\mu}\right] \gamma_5\right\}_{\alpha\beta}\ ,
\end{equation}
where $\xi=\frac{l^+}{M_B}$ is the momentum fraction for the light
spectator quark in the B meson and $\Delta(\xi, \mathbf{l}_{\bot})
=M_B \int_0^{\xi} d\xi' (\Psi^-_B(\xi',\mathbf{l}_{\bot})
-\Psi^+_B(\xi',\mathbf{l}_{\bot}))$. Note the four-component
$l_\perp^\mu$ in Eq.(\ref{projectorB}) is defined through,
$l^{\mu}_\perp=l^\mu-\frac{(l^+ n^\mu +l^-\bar{n}^\mu)}{2}$ with
$l=(\frac{l^+}{\sqrt{2}},\frac{l^-}{\sqrt{2}},\mathbf{l}_\perp)$.

\begin{figure}
\centering
\includegraphics[width=0.6\textwidth]{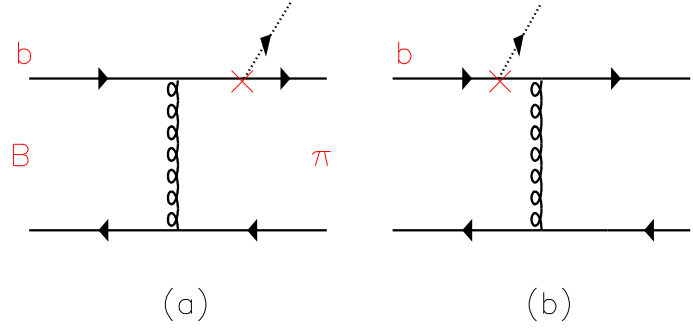}
\caption{Lowest order hard-scattering kernel for $B\to\pi$ form
factor, where the cross denotes an appropriate gamma matrix.}
\label{figbpi}
\end{figure}

In the large recoil regions, the $B\to\pi$ transition form factor
is dominated by a single gluon exchange in the lowest order as
depicted in Fig.(\ref{figbpi}). In the hard scattering kernel, the
transverse momentum in the denominators are retained to regulate
the endpoint singularity. The masses of the light quarks and the
mass difference ($\bar\Lambda$) between the b quark and the B
meson are neglected. The terms proportional to $\mathbf{k}_\bot^2$
or $\mathbf{l}_\bot^2$ in the numerator are dropped, which are
power suppressed compared to other ${\cal O}(M_B^2)$ terms. Under
these treatment, the Sudakov form factor from $k_T$ resummation
can be introduced into the PQCD factorization theorem without
breaking the gauge invariance\cite{li1}. In the transverse
configuration $b$-space and by including the Sudakov form factors
and the threshold resummation effects, we obtain the formulae for
$F_+^{B\pi}(q^2)$ and $F_0^{B\pi}(q^2)$ as following,
\begin{eqnarray}
F_+^{B\pi}(q^2) &=& \frac{\pi C_F}{N_c} f_{\pi}f_BM_B^2\int d\xi
dx\int b_Bdb_B~ b_\pi db_\pi~ \alpha_s(t)
\times\exp(-S(x,\xi,b_\pi,b_B;t)) \nonumber\\
&\times& S_t(x)S_t(\xi)\Bigg \{ \Bigg [ \Psi_\pi(x, b_\pi)\left (
(x\eta+1)\Psi_B(\xi, b_B)+  (x\eta-1)\bar\Psi_B(\xi, b_B) \right
)\nonumber \\
&+& \frac{m_0^p}{M_B}\Psi_p(x, b_\pi)\cdot\left(
(1-2x)\Psi_B(\xi,b_B)+\left(\frac{2}{\eta}-1\right)\bar\Psi_B(\xi,b_B)\right)
-\frac{m_0^p}{M_B}\frac{\Psi'_\sigma(x,b_\pi)}{6} \cdot\nonumber\\
&&\!\!\!\!\!\!\!\!\!\left(\left(1+2x-\frac{2}
{\eta}\right)\Psi_B(\xi,b_B)-\bar\Psi_B(\xi,b_B) \right)
+6\frac{m_0^p}{M_B}\frac{\Psi_\sigma(x,b_\pi)}
{6}\Psi_B(\xi,b_B)\Bigg ]h_1(x,\xi,b_{\pi},b_B)\nonumber \\
&-& (1+\eta+x\eta) \frac{m_0^p}{M_B}
\frac{\Psi_\sigma(x,b_\pi)}{6}
[M_B\Delta(\xi,b_B)]h_2(x,\xi,b_{\pi},b_B)\nonumber \\
&+& \Bigg [ \Psi_\pi(x, b_\pi)\left ( -\xi\bar\eta
[\Psi_B(\xi,b_B)+\bar\Psi_B(\xi,b_B)]+\frac{\Delta(\xi,b_B)}{M_B}
\right )+2\frac{m_0^p}{M_B}\Psi_p(x,b_\pi)\cdot\nonumber \\
&& \left( (1-\xi)\Psi_B(\xi,b_B)+(1+\xi-\frac{2\xi}
{\eta})\bar\Psi_B(\xi,b_B) +2\frac{\Delta(\xi,b_B)} {M_B}\right )
\Bigg ] h_1(\xi,x,b_B,b_\pi) \Bigg \}, \label{fbc+}
\end{eqnarray}
and
\begin{eqnarray}
F_0^{B\pi}(q^2) &=& \frac{\pi C_F}{N_c} f_{\pi}f_BM_B^2\int d\xi
dx\int b_Bdb_B~ b_\pi db_\pi~ \alpha_s(t)
\times \exp(-S(x,\xi,b_\pi,b_B;t))\nonumber\\
&\times& S_t(x)S_t(\xi) \Bigg \{ \Bigg [ \Psi_\pi(x,
b_\pi)\eta\left ( (x\eta+1)\Psi_B(\xi,
b_B)+(x\eta-1)\bar\Psi_B(\xi, b_B) \right )
\nonumber \\
&+&\frac{m_0^p}{M_B}\Psi_p(x, b_\pi) \big((2-\eta-2x\eta)
\Psi_B(\xi,b_B)+\eta\bar\Psi_B(\xi,b_B)\big) \nonumber \\
&-&\frac{m_0^p}{M_B}\frac{\Psi'_\sigma(x,b_\pi)}{6}\cdot\big(
\eta(2x-1)\Psi_B(\xi,b_B)-(2-\eta)\bar\Psi_B(\xi,b_B)
\big) \nonumber\\
&+&6\frac{m_0^p}{M_B}
\eta\frac{\Psi_\sigma(x,b_\pi)}{6}\Psi_B(\xi,b_B)\Bigg ]
h_1(x,\xi,b_\pi,b_B) \nonumber \\
&-& [3-\eta-x\eta]\frac{m_0^p}{M_B} \frac{\Psi_\sigma(x,b_\pi)}{6}
[M_B\Delta(\xi,b_B)]h_2(x,\xi,b_{\pi},b_B)\nonumber \\
&+& \Bigg [ \Psi_\pi(x, b_\pi)\eta\left ( \xi\bar\eta
(\Psi_B(\xi,b_B)+\bar\Psi_B(\xi,b_B))+\frac{\Delta(\xi,b_B)}{M_B}
\right )\nonumber \\
&+& 2\frac{m_0^p}{M_B}\Psi_p(x,b_\pi)\cdot\Big(
(\eta(1+\xi)-2\xi)\Psi_B(\xi,b_B) +\eta(1-\xi)\bar\Psi_B(\xi,b_B)
\nonumber\\
&+& 2(2-\eta) \frac{\Delta(\xi,b_B)}{M_B}\Big) \Bigg ]
h_1(\xi,x,b_B,b_\pi) \Bigg \},\label{fbc0}
\end{eqnarray}
where
\begin{eqnarray}
h_1(x,\xi,b_\pi,b_B)&=&K_0(\sqrt{\xi x\eta}~M_B b_B)
    \Bigg [ \theta(b_B-b_\pi)I_0(\sqrt{x\eta}~M_Bb_\pi)
      K_0(\sqrt{x\eta}~M_B b_B) \nonumber \\
&&+\theta(b_\pi-b_B)I_0(\sqrt{x\eta}~M_Bb_B)
    K_0(\sqrt{x\eta}~M_B b_\pi) \Bigg ], \\
h_2(x,\xi,b_\pi,b_B)&=&\frac{b_B}{2\sqrt{\xi
xy}M_B}K_{1}(\sqrt{\xi x\eta}~M_B b_B)
    \Bigg [ \theta(b_B-b_\pi)I_0(\sqrt{x\eta}~M_Bb_\pi)
      K_0(\sqrt{x\eta}~M_B b_B) \nonumber \\
&&+\theta(b_\pi-b_B)I_0(\sqrt{x\eta}~M_Bb_B)
    K_0(\sqrt{x\eta}~M_B b_\pi) \Bigg ],
\end{eqnarray}
and we have set,
\begin{equation}\label{oldpsi}
\Psi_B=\frac{\Psi_B^{+}+\Psi_B^-}{2}\ ,\;\;\;
\bar{\Psi}_B=\frac{\Psi_B^{+}-\Psi_B^-}{2}\ .
\end{equation}
The functions $I_i$ ($K_i$) are the modified Bessel functions of
the first (second) kind with the $i$-{\it th} order. The angular
integrations in the transverse plane have been performed. The
factor $\exp(-S(x,\xi,b_\pi,b_B;t))$ contains the Sudakov
logarithmic corrections and the renormalization group evolution
effects of both the wave functions and the hard scattering
amplitude,
\begin{equation}
S(x,\xi,b_\pi,b_B;t)=
\left[s(x,b_\pi,M_b)+s(\bar{x},b_\pi,M_b)+s(\xi,b_B,M_b)
-\frac{1}{\beta_{1}}\ln\frac{\hat{t}}{\hat{b}_\pi}
-\frac{1}{\beta_{1}}\ln\frac{\hat{t}}{\hat{b}_B} \right],
\end{equation}
where ${\hat t}={\rm ln}(t/\Lambda_{QCD})$, ${\hat b}_B ={\rm
ln}(1/b_B\Lambda_{QCD})$, ${\hat b}_\pi ={\rm
ln}(1/b_\pi\Lambda_{QCD}) $ and $s(x,b,Q)$ is the Sudakov exponent
factor, whose explicit form up to next-to-leading log
approximation can be found in Ref.\cite{liyu}. $S_t(x)$ and
$S_t(\xi)$ come from the threshold resummation effects and here we
take a simple parametrization proposed in Refs.\cite{li1,kls},
\begin{equation}
S_t(x)=\frac{2^{1+2c}\Gamma(3/2+c)}{\sqrt{\pi}\Gamma(1+c)}
[x(1-x)]^c\;,
\end{equation}
where the parameter $c$ is determined around $0.3$ for the present
case.

The hard scale $t$ in $\alpha_s(t)$ and the Sudakov form factor
might be varied for the different hard scattering parts and here
we need two $t_i$\cite{li1,lucai}, whose values are chose as the
largest scale of the virtualities of internal particles, i.e.
\begin{equation}
t_1={\rm MAX}\left(\sqrt{x\eta}M_B,1/b_\pi,1/b_B\right),\;
t_2={\rm MAX} \left(\sqrt{\xi\eta}M_B,1/b_\pi,1/b_B\right).
\end{equation}
The Fourier transformation for the transverse part of the wave
function is defined as
\begin{equation}\label{fourier}
\Psi(x,\mathbf{b})=\int_{|\mathbf{\mathbf{k}}|<1/b}
d^2\mathbf{k}_\perp\exp\left(-i\mathbf{k}_\perp
\cdot\mathbf{b}\right)\Psi(x,\mathbf{k}_\perp),
\end{equation}
where $\Psi$ stands for $\Psi_\pi$, $\Psi_p$, $\Psi_\sigma$,
$\Psi_B$, $\bar\Psi_B$ and $\Delta$, respectively. The upper edge
of the integration $|\mathbf{k}_\perp|<1/b$ is necessary to ensure
that the wave function is soft enough\cite{huang2}.

In summary, we compare the results in Eqs.(\ref{fbc+},\ref{fbc0})
with those in Refs.\cite{descotes,li1,weiy2,lucai}. In
Ref.\cite{descotes}, only leading twist ($\Psi_\pi$) of the pion
is discussed. Setting the twist-3 terms to zero, the two formulae
in Eqs.(\ref{fbc+},\ref{fbc0}) and Ref.\cite{descotes} are in
agreement. In Ref.\cite{li1}, the single B meson wave function
$\Psi_B$ is assumed and the terms of $\bar\Psi_B$ and $\Delta$ are
neglected. And in Ref.\cite{lucai}, with a new definition for
$\Psi_B$ and $\bar\Psi_B$, i.e.
\begin{equation}
\label{newpsi} \Psi_B=\Psi_B^{+}\ ,\;\;\;
\bar{\Psi}_B=(\Psi_B^{+}-\Psi_B^-),
\end{equation}
both contributions from $\Psi_B$ and $\bar\Psi_B$ are taken into
consideration, with only the terms of $\Delta$ are neglected. The
momentum projector used in \cite{li1,lucai} for the pion is
different from the present projector in Eq.(\ref{benek}), i.e.
there is no term proportional to $\Psi_\sigma$ in
Refs.\cite{li1,lucai}. Except for these
differences\footnote{According to the power counting rules in
Ref.\cite{li1}, the terms that do not existent in Ref.\cite{li1}
are defined as sub-leading terms in $1/M_B$ and are neglected
accordingly. And here, we keep all the terms with care.}, the
formulae in \cite{lucai,li1} are consistent with ours. Our results
agree with Ref.\cite{weiy2}, except for several minus errors that
should be corrected there.

\section{numerical calculations}

In the numerical calculations, we use
\begin{eqnarray}
\Lambda^{(n_f=4)}_{\over{MS}}=250MeV,\;\; f_\pi=131MeV,\;\;
f_B=190MeV,\;\;m^p_{0}=1.30GeV.
\end{eqnarray}
The wavefunctions in the compact parameter $b$-space,
$\Psi^B_+(\xi,b_B)$, $\Psi^B_-(\xi,b_B)$, $\Psi_\pi(x,b_\pi)$,
$\Psi_p(x,b_\pi)$ and $\Psi_\sigma(x,b_\pi)$ can be found in the
appendix. The $k_T$-dependence has been kept in both the B meson
and the pion wavefunctions. As has been argued in several
papers\cite{huangww,weiy,jk,huangwu}, the intrinsic
$k_T$-dependence of the wave function is important and the results
will be overestimated without including this effect, so it is
necessary to include the transverse momentum dependence into the
wave functions not only for the B meson but also for the pion. As
has been argued in Ref.\cite{huangwu}, we take $m^p_0=1.30GeV$ for
latter discussions, which is a little below the value given by the
chiral perturbation theory\cite{chiral}.

\begin{figure}
\centering
\includegraphics[width=0.5\textwidth]{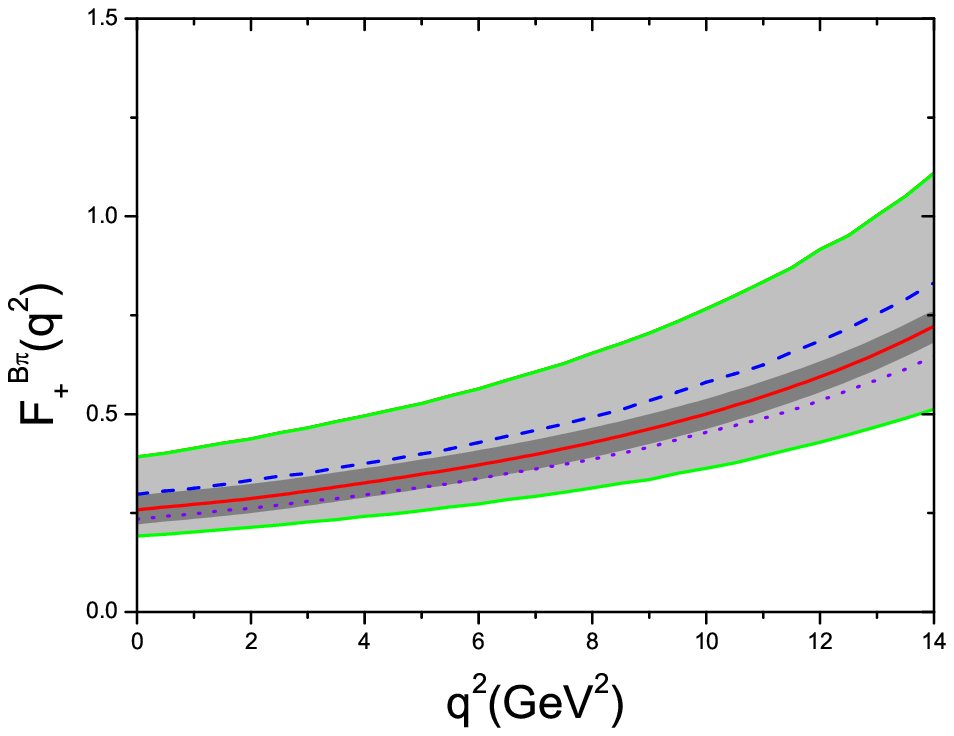}%
\includegraphics[width=0.5\textwidth]{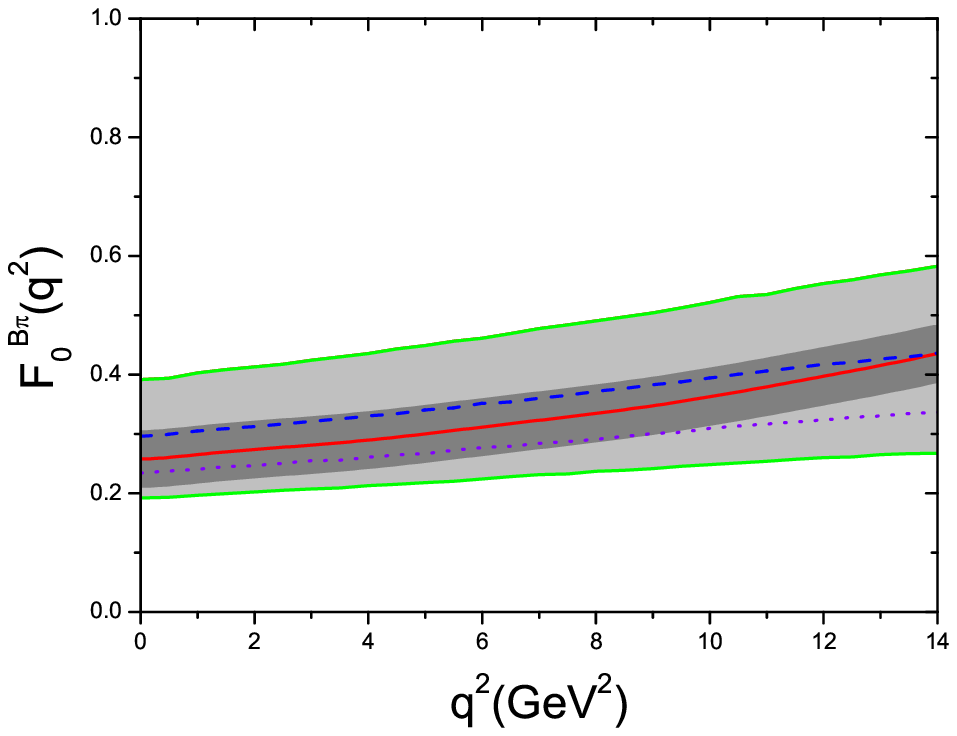}
\caption{PQCD results for the $B\to \pi$ transition form factors
$F_+^{B\pi}(q^2)$ (Left) and $F_0^{B\pi}(q^2)$ (Right) with
different values for $\bar\Lambda$. The dashed line stands for
$\bar\Lambda=0.5GeV$, the dotted line stands for
$\bar\Lambda=0.6GeV$, the upper edge of the shaded band
corresponds to $\bar\Lambda=0.40GeV$ and the lower edge of the
band corresponds to $\bar\Lambda=0.70GeV$. For comparison, the
solid line comes from the QCD LCSR\cite{sumhuang,sumrule} and the
fuscous shaded band shows its theoretical error $\pm 10\%$.}
\label{lambda}
\end{figure}

The two wavefunctions $\Psi_B$ and $\bar\Psi_B$ of the B meson
shown in the appendix depend only on the effective mass
($\bar\Lambda=M_B-m_b$) of the B meson. An estimate of
$\bar\Lambda$ using QCD sum rule approach gives
$\bar\Lambda=0.57\pm0.07GeV$\cite{lambdavalue}. In
Fig.(\ref{lambda}), we show the $B\to\pi$ transition form factor
with different value of $\bar\Lambda$, where the shaded band is
drawn with a broader range for $\bar\Lambda$, i.e.
$\bar\Lambda\in(0.4GeV,0.7GeV)$. And for comparison, we show the
QCD LCSR result \cite{sumrule} in solid line and its theoretical
error $(\pm 10\%)$ by a fuscous shaded band in Fig.(\ref{lambda}).
The results show that the $B\to\pi$ transition form factor will
decrease with the increment of $\bar\Lambda$. When $\bar\Lambda\in
(0.5GeV,0.6GeV)$, one may observe that the present results agree
well with the QCD LCSR results\cite{sumhuang,sumrule} up to
$q^2\sim 14GeV^2$. In Fig.(\ref{lambda}), for simplicity, only the
QCD LCSR results of Ref.\cite{sumrule} are shown. The LCSR results
in Refs.\cite{sumhuang,sumrule} are in agreement with each other
even though they have taken different ways to improve the QCD LCSR
calculation precision, i.e. in Ref.\cite{sumhuang}, an alternative
way to do the QCD LCSR calculation is adopted in which the pionic
twist-3 contributions are avoided by calculating the correlator
with a proper chiral current and then the leading twist
contributions are calculated up to next-to-leading order; while in
Ref.\cite{sumrule}, the usual QCD LCSR approach is adopted and
both the twist-2 and twist-3 contributions are calculated up to
next-to-leading order. In Ref.\cite{lucai}, $\bar\Lambda$ is
treated as a free parameter and a bigger value is adopted there,
i.e. $\bar\Lambda=(0.70\pm0.05)GeV$. The main reason is that in
the present paper, we have used an improved form (with better
endpoint behavior than that of the asymptotic one) for the pionic
twist-3 wavefunction $\Psi_p$, while in Ref.\cite{lucai}, they
took $\phi_p$ in Ref.\cite{pball} (with an endpoint behavior even
worse than the asymptotic one) other than $\Psi_p$ to do the
calculations, so the value of $\bar\Lambda$ in Ref.\cite{lucai}
must be big enough to suppress the endpoint singularity coming
from the hard kernel. For clarity, if not specially stated, we
shall fix $\bar\Lambda$ to be $0.5GeV$ in the following
discussions.

\begin{figure}
\centering
\begin{minipage}[c]{0.48\textwidth}
\centering
\includegraphics[width=2.9in]{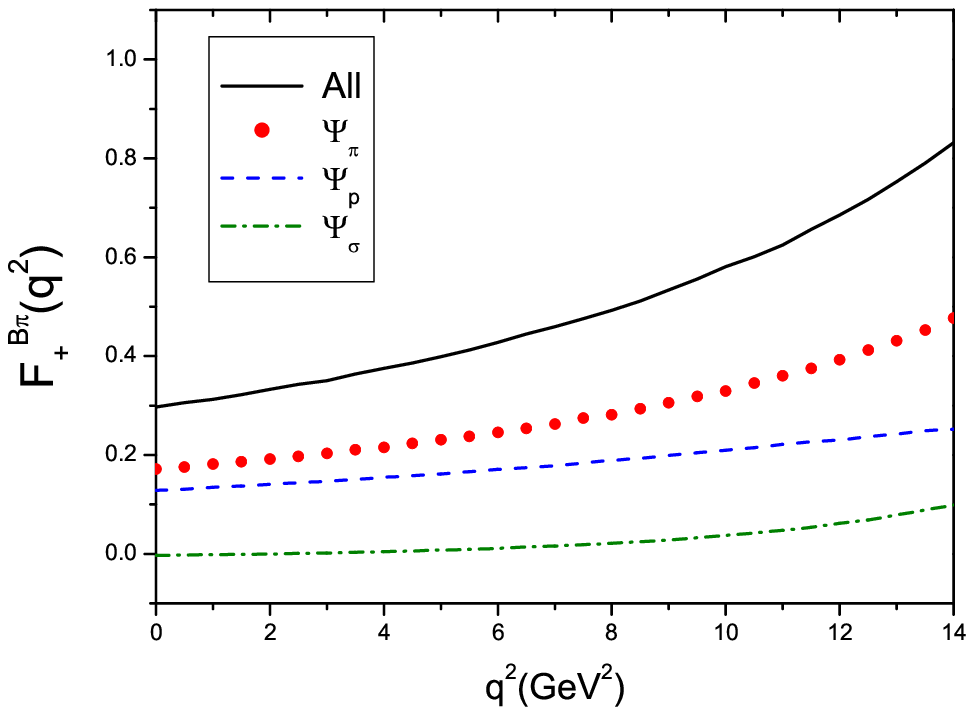}
(a)
\end{minipage}%
\begin{minipage}[c]{0.48\textwidth}
\centering
\includegraphics[width=2.9in]{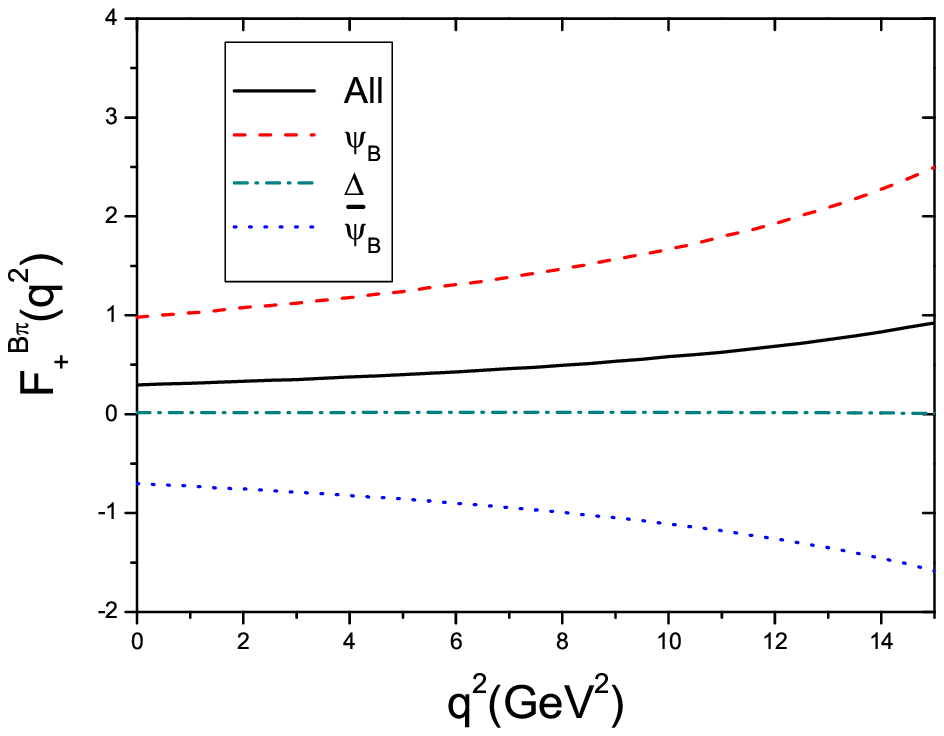}
(b)
\end{minipage}
\caption{PQCD results for the $B\to \pi$ transition form factor
$F_+^{B\pi}(q^2)$ with fixed $\bar\Lambda=0.5GeV$. The left
diagram is for the different pion twist structures, $\Psi_\pi$,
$\Psi_p$ and $\Psi_\sigma$. The right diagram is for the different
B meson structures, $\Psi_B$, $\bar\Psi_B$ and $\Delta$, where
$\Psi_B$ and $\bar\Psi_B$ are defined in Eq.(\ref{oldpsi}).}
\label{pionwave}
\end{figure}

Second, to get a deep understanding of the $B\to\pi$ transition
form factor, we discuss the contributions from different parts of
the B meson wavefunction or the pion wave function,
correspondingly. Here we take $F_+^{B\pi}(q^2)$ to do our
discussions and the case of $F_0^{B\pi}(q^2)$ can be done in a
similar way. In Fig.(\ref{pionwave}a), we show the contributions
from the different twist structures of the pion wave function,
i.e. $\Psi_\pi$, $\Psi_p$ and $\Psi_\sigma$ (the contributions
from the terms involving $\Psi'_\sigma$ are included in
$\Psi_\sigma$), respectively. From Fig.(\ref{pionwave}a), one may
observe that the contribution from $\Psi_\pi$ is the biggest, then
comes that of $\Psi_p$ and $\Psi_\sigma$. And the ratio between
all the twist-3 contributions and the leading twist contribution
is $\sim 70\%$ in the large recoil regions. This behavior is quite
different from the conclusion that has been drawn in
Refs.\cite{weiy2,li1}, in which they concluded that the twist-3
contribution is bigger than that of twist-2 contribution,
especially in Ref.\cite{weiy2}, it claimed that the twist-3
contribution is about three times bigger than that of twist-2 at
$q^2=0$. Such kind of big twist-3 contributions are due to the
fact that they only took the pion distribution amplitudes into
consideration (or simply adding a harmonic transverse momentum
dependence for the pion wavefunctions), and then the endpoint
singularity coming from the hard kernel can not be effectively
suppressed, especially for $\Psi_p$ whose DA's asymptotic behavior
is $\phi_p\equiv 1$. In Ref.\cite{huangwu}, the authors have made
a detailed analysis on the model dependence of the twist-3
contributions to the pion electro-magnetic form factor, and have
raised a new twist-3 wavefunction (as is shown in the appendix)
with a better endpoint behavior for $\Psi_p$, which is inspired
from QCD sum rule moment calculation. With this model wave
function for $\Psi_p$, Ref.\cite{huangwu} shows that the twist-3
contributions of the pion electro-magnetic form factor agree well
with the power counting rule, i.e. the twist-3 contribution drops
fast and it becomes less than the twist-2 contribution at $Q^2\sim
10GeV^2$. For the present B meson case, according to the power
counting rules in Ref.\cite{li1}, the twist-3 contribution and the
twist-2 contribution are of the same order, however one may find
from Fig.(\ref{pionwave}a) that with a new form with better
endpoint behavior for $\Psi_p$, the twist-3 contribution can be
effectively suppressed and then its contribution is less than the
leading twist contribution.

\begin{figure}
\centering
\begin{minipage}[c]{0.48\textwidth}
\centering
\includegraphics[width=2.9in]{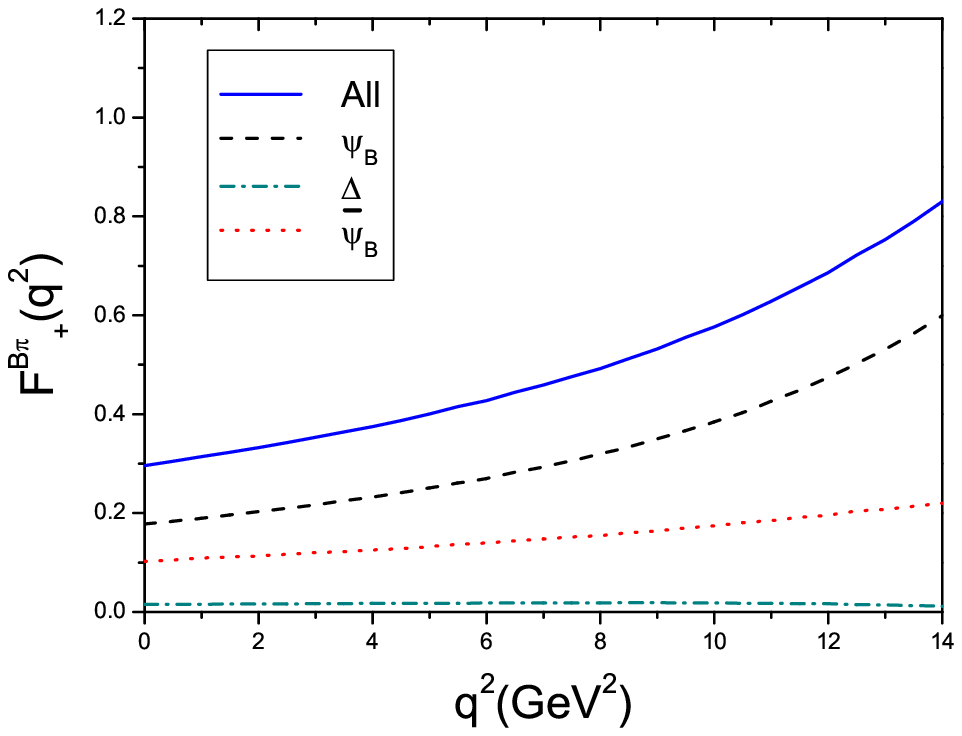}
(a)
\end{minipage}%
\begin{minipage}[c]{0.48\textwidth}
\centering
\includegraphics[width=2.9in]{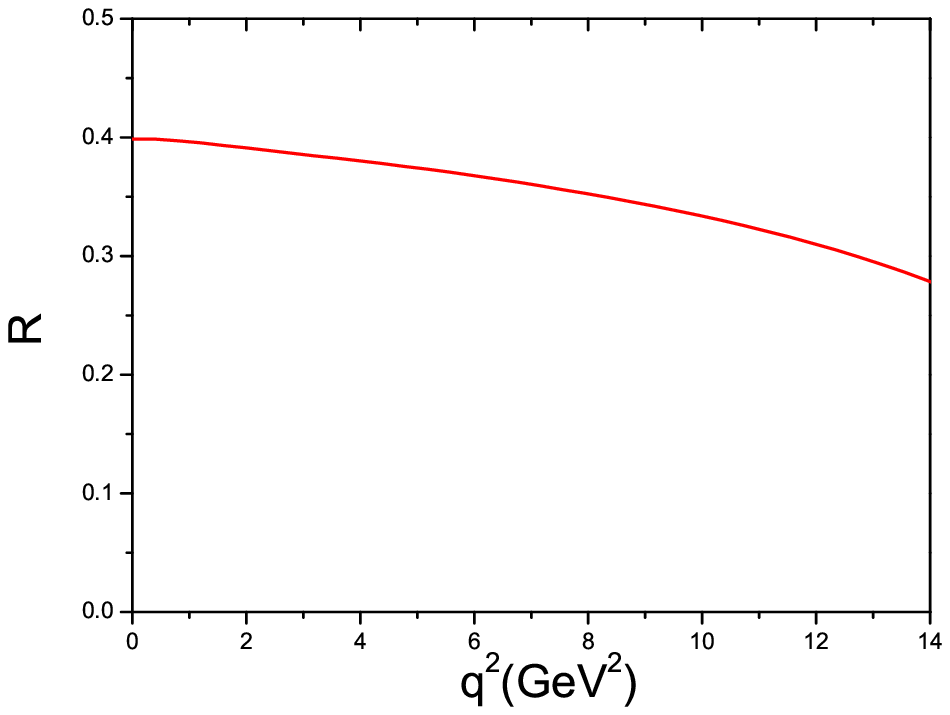}
(b)
\end{minipage}
\caption{PQCD results for the $B\to \pi$ form factor
$F_+^{B\pi}(q^2)$ with fixed $\bar\Lambda=0.5GeV$, where $\Psi_B$
and $\bar\Psi_{B}$ are defined in Eq.(\ref{newpsi}). The left
diagram shows the contributions from different B meson
wavefunctions, $\Psi_B$, $\bar\Psi_{B}$ and $\Delta$,
respectively. The right diagram is the distribution of the ratio
$R=\left(\frac{F^{B\pi}_{+}|_{\bar\Psi_B}}{F^{B\pi}_{+}|_{All}}\right)$
versus $q^2$. } \label{bwaves2}
\end{figure}

Now, we show to what extent, $\bar\Psi_B$ will affect the final
results. Fig.(\ref{pionwave}b) presents the contributions from
$\Psi_B$, $\bar\Psi_B$ and $\Delta$ respectively, where $\Psi_B$
and $\bar\Psi_{B}$ are defined in Eq.(\ref{oldpsi}). From
Fig.(\ref{pionwave}b), one may observe that the contribution from
$\Delta$ is quite small and can be safely neglected as has been
done in most of the calculations. However the contribution from
$\bar\Psi_B$ is quite large, i.e. at $q^2=0$, the ratio between
the contributions of $\bar\Psi_B$ and $\Psi_B$ is about $(-70\%)$,
which roughly agrees with the observation in Ref.\cite{weiy2}. So
the negative contribution from $\bar\Psi_B$ can not be neglected,
and it is necessary to suppress the big positive contribution from
$\Psi_B$ so as to get a more reasonable total contributions from
both $\Psi_B$ and $\bar\Psi_B$. The above results of
Fig.(\ref{pionwave}b) is obtained by using the definition
Eq.(\ref{oldpsi}). A new definition (\ref{newpsi}) for $\Psi_B$
and $\bar\Psi_B$ has been raised in Ref.\cite{lucai} and the
contributions from the $\Psi_B$, $\bar\Psi_B$ and $\Delta$ with
such a new definition (\ref{newpsi}) are shown in
Fig.(\ref{bwaves2}a). We draw the distribution of the
corresponding ratio $R=\left(\frac{F^{B\pi}_{+}
|_{\bar\Psi_B}}{F^{B\pi}_{+}|_{All}}\right)$ versus $q^2$ in
Fig.(\ref{bwaves2}b), where $\left(F^{B\pi}_{+}
|_{\bar\Psi_B}\right)$ means that only the contributions from
$\bar\Psi_B$ are considered and $\left(F^{B\pi}_{+}|_{All}\right)$
means that all the contributions from the B meson wavefunctions
are taken into consideration. One may observe from
Fig.(\ref{bwaves2}b) that even with the new definition
(\ref{newpsi}) for $\Psi_B$ and $\bar\Psi_B$, the contribution
from $\bar\Psi_B$ is not small ($\sim 25-40\%$) and it can not be
safely neglected. Thus both $\Psi_B$ and $\bar\Psi_B$ should be
kept in the calculation for giving a better understanding of the B
decays.

\begin{figure}
\centering
\includegraphics[width=0.5\textwidth]{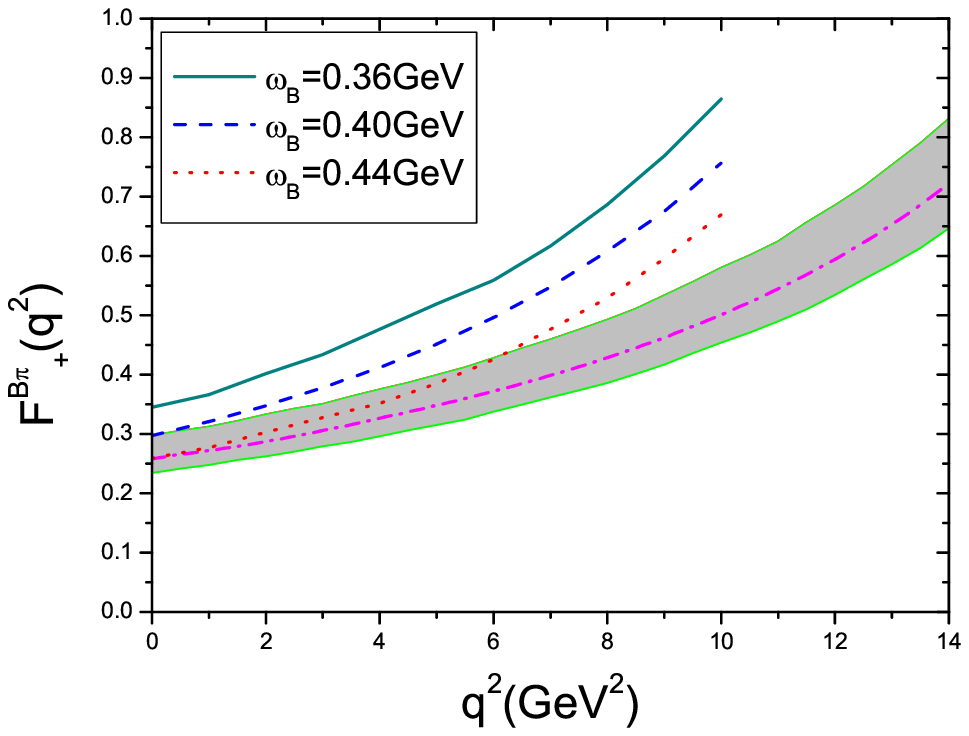}%
\includegraphics[width=0.5\textwidth]{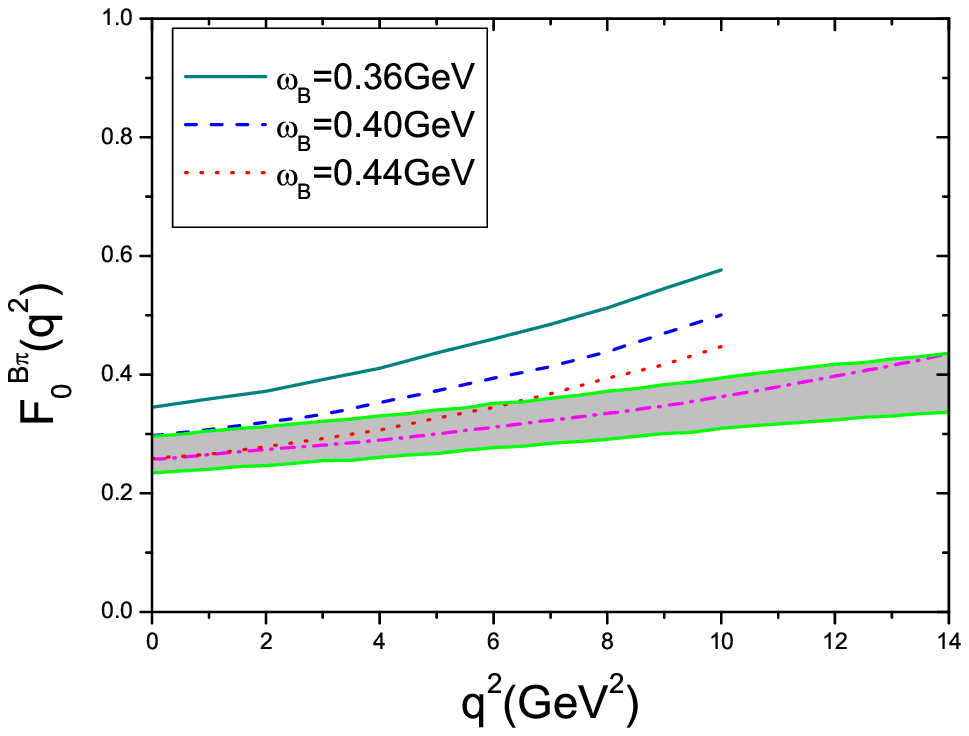}
\caption{Comparison of different PQCD results for the $B\to \pi$
transition form factors $F_+^{B\pi}(q^2)$ (Left) and
$F_0^{B\pi}(q^2)$ (Right). The solid, dashed and dotted lines are
the results obtained in Ref.\cite{li1} and are for
$\omega_B=0.36$GeV, $0.40$GeV and $0.44$GeV respectively. The
shaded band are our present results with the upper edge for
$\bar\Lambda=0.50GeV$ and the lower edge for
$\bar\Lambda=0.60GeV$, respectively. For comparison, the dash-dot
line stands from the QCD LCSR result\cite{sumhuang,sumrule}.}
\label{lambda2}
\end{figure}

Finally, we make a comparison of the present results for
$F_{+,0}^{B\pi}(q^2)$ with those obtained in Ref.\cite{li1} in
Fig.(\ref{lambda2}). In Ref.\cite{li1}, $\bar\Psi_B$ has been
neglected and $\Psi_B$ takes the form
\begin{equation}
\Psi_B(x,b_B)=N_Bx^2(1-x)^2\exp\left[-\frac{1}{2}\left(
\frac{xM_B}{\omega_B}\right)^2-\frac{\omega_B^2 b_B^2}{2}\right],
\end{equation}
where $N_B$ is the normalization factor and $\omega_B$ is taken to
be $(0.40\pm 0.04)GeV$. In Fig.(\ref{lambda2}), we show their
results for $\omega_B=0.36$GeV, $0.40$GeV and $0.44$GeV and our
present results with $\bar\Lambda\in (0.5GeV, 0.6GeV)$,
respectively. The two results in the large recoil regions $q^2\sim
0$ are consistent with each other, however one may observe that
the fast rise in Ref.\cite{li1} has been suppressed in our present
results and the slope of the present obtained form factors
$F_{+,\;0}^{B\pi}(q^2)$ are more consistent with the QCD LCSR
results in Ref.\cite{sumhuang,sumrule}. The main reason for the
differences between our present results and those in
Ref.\cite{li1} is that we have used a better endpoint behavior
wavefunction for $\Psi_p$\cite{huangwu}. With this new form for
$\Psi_p$, we find that the total twist-3 contributions are in fact
less than ($\sim 70\%$) the leading twist contribution in the
large recoil regions. While in Ref.\cite{li1}, the twist-3
contributions are about two times bigger than that of the leading
twist, especially for the bigger $q^2$ regions, and then the total
contributions will give a fast rise in shape.

\section{discussion and summary}

In the present paper, we have examined the $B\to\pi$ transition
form factor in the PQCD approach, where the transverse momentum
dependence for the wavefunction, the Sudakov effects and the
threshold effects are included to regulate the endpoint
singularity and to derive a more reasonable result. We emphasize
that the transverse momentum dependence for both the B meson and
the pion is important to give a better understanding of the
$B\to\pi$ transition form factor. The pionic twist-3 contributions
to the $B\to\pi$ transition form factor are carefully studied with
a better endpoint behavior wavefunction for $\Psi_p$, and
Fig.(\ref{pionwave}) shows that the twist-3 contributions are of
the same order of the leading twist contribution, however its
values are less than that of the leading twist. This observation
improves the results obtained in Refs.\cite{weiy2,li1}, in which
the asymptotic behavior for $\phi_p$ was used and they claimed a
large twist-3 contributions to the $B\to\pi$ transition form
factor, i.e. bigger than that of the leading twist.
Fig.(\ref{pionwave}b) and Fig.(\ref{bwaves2}) show that both
$\Psi_B$ and $\bar\Psi_B$ are important, no matter what definition
(Eq.(\ref{oldpsi}) or Eq.(\ref{newpsi})) is chosen. Under the
definition (\ref{oldpsi}), the negative contribution from
$\bar\Psi_B$ is necessary to suppress the big contribution from
$\Psi_B$ and to obtain a reasonable total contributions. While
under the definition Eq.(\ref{newpsi}), the contribution from
$\bar\Psi_B$ is power suppressed to that of $\Psi_B$, however it
still can contribute $25-40\%$ to the total contributions. As is
shown in Fig.(\ref{lambda2}), a comparison of our present results
for $F_{+,\;0}^{B\pi}(q^2)$ with those in Ref.\cite{li1} shows
that a better PQCD result (with its slope closes to the QCD LCSR
results) can be obtained by carefully considering both the pionic
twist-3 contributions and the contributions from the two
wavefunctions $\Psi_B$ and $\bar\Psi_B$ of the B meson.

\begin{figure}
\centering
\includegraphics[width=0.5\textwidth]{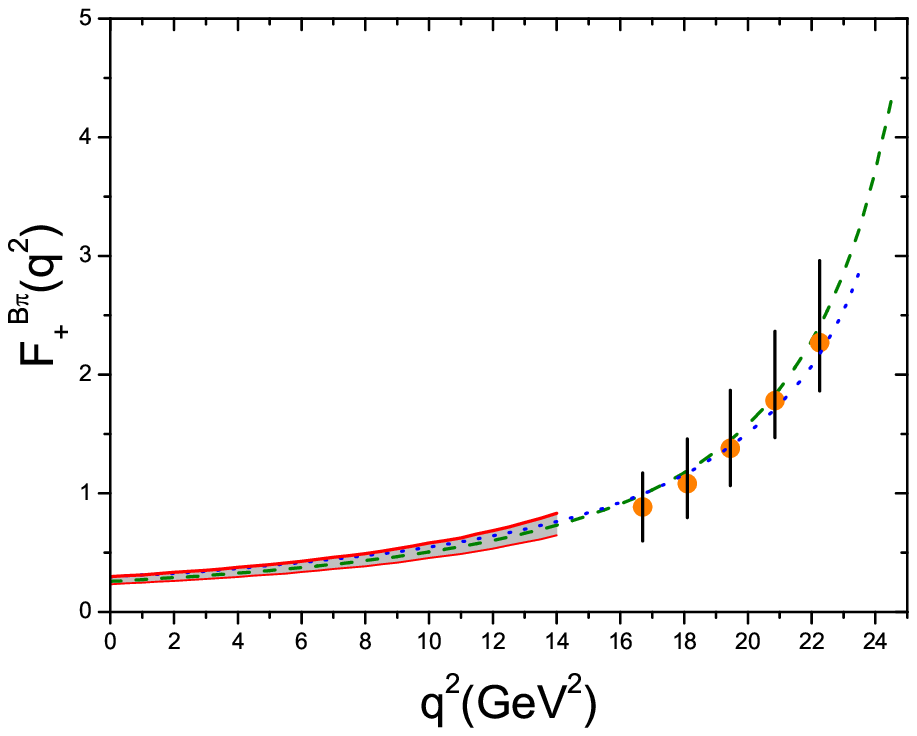}%
\includegraphics[width=0.5\textwidth]{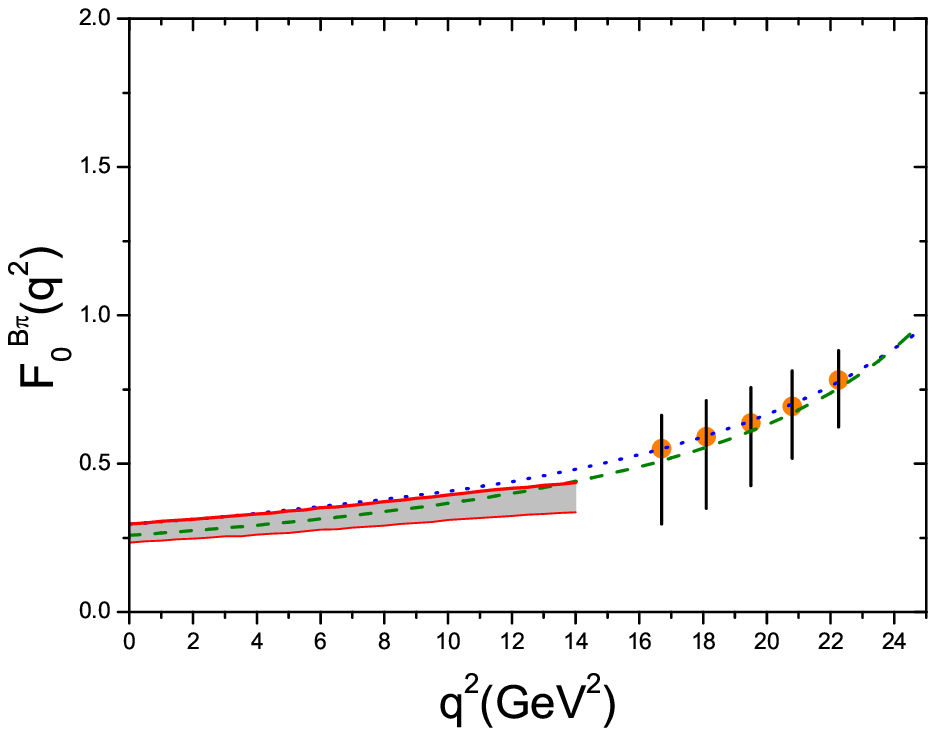}
\caption{PQCD results for the $B\to \pi$ form factors
$F_+^{B\pi}(q^2)$ (Left) and $F_0^{B\pi}(q^2)$ (Right). The shaded
band are our present results with the upper edge for
$\bar\Lambda=0.50GeV$ and the lower edge for
$\bar\Lambda=0.60GeV$, respectively. The dashed and dotted lines
stand for the QCD LCSR result Eq.(\ref{sumruleapp}) and the fits
to the lattice QCD results with errors\cite{lattice3},
respectively. } \label{combined}
\end{figure}

In the literature, the values of the $B\to\pi$ transition form
factors $F^{B\pi}_{+}(0)$ and $F^{B\pi}_{0}(0)$ are determined
around $0.3$. With $\bar\Lambda\in (0.50GeV,0.60GeV)$, we obtain
$F^{B\pi}_{+,0}(0)=0.265\pm 0.032$. This result is consistent with
the extrapolated lattice QCD result $F^{B\pi}_{+,0}(0)=0.27\pm
0.11$\cite{lattice} and the newly obtained QCD LCSR result
$F^{B\pi}_{+,0}(0)=0.258\pm0.031$\cite{sumrule}. The PQCD
calculation are reliable only when the involved energy scale is
hard enough. The lattice QCD calculations which presently are
available only for the soft regions, i.e. $q^2>15GeV^2$. The QCD
LCSR can treat both hard and soft contributions with $q^2\precsim
18GeV^2$\cite{sumhuang,sumrule} on the same footing. Therefore,
the results from the PQCD approach, the lattice QCD approach and
the QCD LCSRs are complementary to each other and by combining the
results of those three approaches, one may obtain an understanding
of the $B\to\pi$ transition form factor in the whole physical
regions. The $B\to\pi$ transition form factors $F^{B\pi}_+(q^2)$
and $F^{B\pi}_0(q^2)$ derived from QCD LCSRs can be written in the
following parameterization \cite{sumrule}:
\begin{equation}\label{sumruleapp}
F^{B\pi}_+(q^2)=\frac{r_1}{1-q^2/m_1^2} + \frac{r_2}{1-q^2/m_{\rm
fit}^2},\;\;\;F^{B\pi}_0(q^2)=\frac{r_3}{1-q^2/m_{\rm 0fit}^2},
\end{equation}
where $r_1$, $r_2$, $r_3$, $m_1$, $m_{\rm fit}$ and $m_{\rm 0fit}$
are fitted parameters and can be taken as\cite{sumrule},
$r_1=0.744$, $r_2=-0.486$, $r_3=0.258$, $m_1=5.32GeV$, $m_{\rm
fit}^2=40.73GeV^2$ and $m_{\rm 0fit}^2=33.81GeV^2$. With the
parameterization Eq.(\ref{sumruleapp}), the QCD LCSR results can
be extrapolated up to the upper limit of $q^2$, i.e. $q^2\sim
25GeV^2$, and then it can be treated as a bridge to connect both
the PQCD results and the lattice QCD results. In
Fig.(\ref{combined}), we show the results of the PQCD approach,
the lattice QCD approach and the extrapolated QCD LCSR results
defined in Eq.(\ref{sumruleapp}), respectively. Our present PQCD
results with $\bar\Lambda\in (0.5GeV, 0.6GeV)$ are in agreement
and can match with the QCD LCSR results and the lattice QCD
calculations, which are shown in Fig.(\ref{combined}).

In summary, we have shown that the PQCD approach can be applied to
calculate the $B\to\pi$ transition form factor in the large recoil
regions. The twist-3 contributions are less than the leading twist
contribution with a better endpoint behavior twist-3 wavefunctions
and both of the two wavefunctions $\Psi_B$ and $\bar\Psi_B$ of the
B meson are necessary to give a deep understanding of the B
decays, e.g. $B\to\pi$ transition form factor. Combining the PQCD
results with the QCD LCSR and the lattice QCD calculations, the
$B\to\pi$ transition form factor can be determined in the whole
kinematic regions.

\begin{center}
\section*{Acknowledgements}
\end{center}

The authors would like to thank H.N. Li for some useful
discussions. This work was supported in part by the Natural
Science Foundation of China (NSFC). \\

\appendix
\section{formulae for the pion and B meson wavefunctions}

To do the numerical calculations, for the pion wave functions, we
take
\begin{equation}
\Psi_{\pi,\sigma}(x,\mathbf{k}_{\perp})= A_{\pi} \exp
\left(-\frac{m^2+k_\perp^2}{8\beta^2 x(1-x)} \right),
\end{equation}
where the parameters can be determined by the normalization
condition of the wave function
\begin{equation}\label{normalization}
\int^1_0 dx \int \frac{d^{2}{\bf k}_{\perp}}{16\pi^3}\Psi(x,{\bf
k}_{\perp}) =1\ ,
\end{equation}
and some necessary constraints\cite{hms}. And one can construct a
model wave function $\Psi_p$ with $k_T$ dependence in the
following\cite{huangwu},
\begin{equation}\label{hel0wave}
\Psi_p(x,\mathbf{k}_{\perp}) = (1+B_p C^{1/2}_2(1-2x)+C_p
C^{1/2}_4(1-2x))\frac{A_p}{x(1-x)}\exp \left(
-\frac{m^2+k_\perp^2}{8\beta^2x(1-x)}\right),
\end{equation}
where $C^{1/2}_2(1-2x)$ and $C^{1/2}_4(1-2x)$ are Gegenbauer
polynomials and the coefficients $A_p$, $B_p$ and $C_p$ can be
determined by the DA moments. In the above equations,
\begin{equation}\label{parameter}
m=290MeV,\;\;\beta=385MeV,
\end{equation}
which are derived for $\langle\mathbf{k_\perp}^2\rangle\approx
(356MeV)^2$\cite{hms}. The parameters in Eq.(\ref{hel0wave}) can
then be determined as,
\begin{equation}
A_{\pi}=1.187\times 10^{-3}MeV^{-2},\;\;\;A_p=2.841\times
10^{-4}MeV^{-2},\;\;\; B_p=1.302,\;\;\; C_p=0.126.
\end{equation}

And for the B meson wave function, we take\cite{qiao,wu1}
\begin{eqnarray}
\Psi^-_B(\xi,\mathbf{k}_{\perp})&=&16\pi^3\frac{2\bar\xi-\xi}{2\pi\bar\xi^2}
\theta(2\bar\xi-\xi)\delta(k_\perp^2-M_B^2\xi(2\bar\xi-\xi)),\\
\Psi^+_B(\xi,\mathbf{k}_{\perp})&=&16\pi^3\frac{\xi}{2\pi\bar\xi^2}\theta(2\bar\xi-\xi)
\delta(k_\perp^2-M_B^2\xi(2\bar\xi-\xi)),
\end{eqnarray}
with $\xi=\frac{l^+}{M_B}$ and $\bar\xi=\frac{\bar\Lambda}{M_B}$,
where $\bar\Lambda$ is the effective mass of the B meson.

After doing the Fourier transformation with the formula
Eq.(\ref{fourier}), we obtain
\begin{eqnarray}
\Psi_{\pi,\sigma}(x,b_\pi)&=&2\pi A_{\pi}\int_0^{1/b_\pi}
\exp\left(-\frac{m^2}{8\beta^2x(1-x)}\right)J_0(b_\pi k_\perp)
k_\perp dk_\perp\\
\Psi_p(x,b_\pi)&=&\frac{2\pi A_p}{x(1-x)}[1+B_p
C^{1/2}_2(1-2x)+C_pC^{1/2}_4(1-2x)]\cdot\nonumber\\
&&\int_0^{1/b_\pi} \exp\left(-\frac{m^2}
{8\beta^2x(1-x)}\right)J_0(b_\pi k_\perp)k_\perp dk_\perp\\
\Psi^-_B(\xi,b_B)&=&16\pi^3\frac{2\bar\xi-\xi}{2\bar\xi^2}
\theta(2\bar\xi-\xi)\theta(1/b_B^2-\xi(2\bar{\xi}-\xi
)M_B^2)J_0(M_B b_B\sqrt{\xi(2\bar{\xi}-\xi)})\\
\Psi^+_B(\xi,b_B)&=&16\pi^3\frac{\xi}{2\bar\xi^2}
\theta(2\bar\xi-\xi)\theta(1/b_B^2-\xi(2\bar{\xi}-\xi
)M_B^2)J_0(M_B b_B\sqrt{\xi(2\bar{\xi}-\xi)})\\
\Delta(\xi,b_B)&=&M_B\int^\xi_0 d\xi' [\Psi^-_B(\xi',b_B)-
\Psi^+_B(\xi',b_B)]\nonumber\\
&=&16\pi^3 M_B\theta(1/b_B^2-\xi(2\bar{\xi}-\xi
)M_B^2)\int^{\xi}_0\frac{\bar\xi-\xi'}{\bar\xi^2} J_0\left(M_B
b_B\sqrt{\xi'(2\bar{\xi}-\xi')}\right)d\xi'.
\end{eqnarray}
One may easily find that the effects of the upper limit
($1/b_{B}$) for the B meson wave functions are quite small
(numerically less than $0.1\%$). This is reasonable, since the B
meson mass is enough heavy to give a natural separation scale.

\end{document}